\documentclass[prl,showpacs,preprint,amsmath,amssymb,groupadress,superscriptaddress]{revtex4}
\usepackage{graphicx}
\begin{document}

\title{Ferroelectricity induced by acentric spin-density waves in YMn$_2$O$_5$}

\author{L.C. Chapon}
\affiliation{ISIS facility, Rutherford Appleton Laboratory-CCLRC,
Chilton, Didcot, Oxfordshire, OX11 0QX, United Kingdom. }
\author{P.G. Radaelli}
\affiliation{ISIS facility, Rutherford Appleton Laboratory-CCLRC,
Chilton, Didcot, Oxfordshire, OX11 0QX, United Kingdom. }
\affiliation{Dept. of Physics and Astronomy, University College
London, Gower Street, London WC1E 6BT, United Kingdom}
\author{G.R. Blake}
\affiliation{ISIS facility, Rutherford Appleton Laboratory-CCLRC,
Chilton, Didcot, Oxfordshire, OX11 0QX, United Kingdom. }
\affiliation{Materials Science Division, Argonne National
Laboratory, Argonne, Illinois 60439, USA}
\author{S. Park}
\affiliation{Department of Physics and Astronomy, Rutgers
University, Piscataway, New Jersey 08854, USA}
\author{S-W. Cheong}
\affiliation{Department of Physics and Astronomy, Rutgers
University, Piscataway, New Jersey 08854, USA}

\date{\today}

\begin{abstract}
The commensurate and incommensurate magnetic structures of the
magnetoelectric system YMn$_{2}$O$_{5}$, as determined from neutron
diffraction, were found to be spin-density waves lacking a global
center of symmetry. We propose a model, based on a simple
magneto-elastic coupling to the lattice, which enables us to predict
the polarization based entirely on the observed magnetic structure.
Our data accurately reproduce the temperature-dependence of the
spontaneous polarization, in particular its sign reversal at the
commensurate-incommensurate transition.
\end{abstract}


\maketitle

There is currently great interest in understanding the microscopic
nature of the coupling between ferroelectricity and magnetic
ordering in several transition metal oxides, such as RMnO$_{3}$ and
RMn$_{2}$O$_{5}$ (R=rare earth
element)\cite{Kimura,Hur,chapon:177402,blake:214402,kenzelmann:087206,ratcliff:060407}.
This coupling is responsible for the sensitivity of these materials
to an applied magnetic field and may lead to new classes of
functional materials. Unlike more conventional multiferroics such as
BiFeO$_{3}$ and BiMnO$_{3}$, the paramagnetic phase in these new
materials is centrosymmetric, and electrical polarization appears
only at the transition to a magnetically ordered phase. This implies
that the ordered spin structure is responsible for removing the
center of symmetry and generating a polar field. Two approaches have
so far been proposed in the literature:  the magneto-elastic effect
could occur through a \emph{scalar} field of the type $S_{n} \cdot
S_{n+1}$, which must be coupled to a pre-existing polar field from
the crystal structure, or through a \emph{vector} field of the type
$S_{n} \times S_{n+1}$\cite{Dagotto,ratcliff:060407,lawes:087205}.
In the latter case, non-collinearity is a key ingredient to promote
a polar state, whereas in the former case a collinear phase could in
principle support electrical polarization.  Naturally, in both
cases, global inversion symmetry must be lost.

Recently we showed that, for TbMn$_{2}$O$_{5}$\cite{chapon:177402},
the largest electric polarization is associated with a commensurate
magnetic (CM) state that is almost collinear.  The magnetic
structure can be described as a superposition of several
amplitude-modulated waves on inequivalent lattice sites with
non-coincident nodal points, making it acentric (the structure has
constant moments for an appropriate choice of the global phase). On
further cooling below 25 K, the TbMn$_{2}$O$_{5}$ magnetic structure
becomes incommensurate (ICM) with \textbf{k}$\sim$(0.48,0,0.32).
Although the electrical polarization evolves in a complex way
through this transition, the ICM phase remains ferroelectric,
displaying, at low temperatures, a remarkably strong coupling with
an applied magnetic field\cite{Hur}.  It is therefore of great
interest to solve the ICM structure and determine how global
inversion symmetry is lost, since in a simple spin density wave
(SDW) one can always find a lattice point that is also an inversion
center. On the basis of theoretical considerations and experimental
data, Kenzelmann and coworkers\cite{kenzelmann:087206} propose that
in TbMnO$_{3}$ inversion symmetry is broken by the development of a
cycloidal magnetic structure, which can be described as a
superposition in quadrature of two waves associated with different
components of the magnetic moment on the \emph{same} site.  Here, we
propose a different and, to our knowledge, hitherto unexplored
mechanism for the ICM phases of the \textit{R}Mn$_{2}$O$_{5}$
series: the loss of inversion symmetry arises from the superposition
of two waves on \emph{different} crystallographic sites, each with
an independent phase factor. In the specific case of
YMn$_{2}$O$_{5}$, where we have solved both CM and ICM structures
from neutron diffraction data, we show that the temperature
dependence of the electrical polarization as calculated from the
magnetic structure using a simple $S_{n} \cdot S_{n+1}$ exchange
coupling is consistent with the measurement of electrical
properties\cite{Kagomiya2}. In particular, our model is capable of
reproducing the sign reversal of the polarization observed at the
CM-ICM transition\cite{Kagomiya2}.

YMn$_{2}$O$_{5}$ is isostructural to TbMn$_{2}$O$_{5}$ and shows the
same sequence of magnetic transitions and electrical anomalies upon
cooling, but the analysis of the magnetic structures is considerably
simplified by the absence of magnetism on the rare earth site.
Magnetic ordering appears below 45K with a CM vector, and switches
to an ICM state below 23K. Similar to TbMn$_{2}$O$_{5}$, the
ferroelectric state coexists with the magnetically ordered state and
at the first-order CM-ICM transition, the dielectric constant jumps
to higher values whereas the spontaneous electrical polarization is
reversed and decreases in amplitude to about 25\%\ of its original
value\cite{Kagomiya2}. Polycrystalline YMn$_{2}$O$_{5}$ was prepared
by conventional solid-state reaction in an oxygen environment.
Neutron powder diffraction data were collected using the GEM
diffractometer at the ISIS facility. Data were recorded on warming
from 1.9K to 53K in 2K steps using a helium cryostat. A collection
time of 2 hours was used at 1.9K in order to obtain high statistics
data in the saturated ICM regime and 20 minutes for all other
temperatures. Data analysis was performed with the program
FullProF\cite{Fullprof}. Magnetic structures were determined by
using global optimization techniques (Simulated Annealing) for data
collected at respectively 1.9K and 24.7K followed by final Rietveld
refinements at all temperatures. During simulated annealing runs,
the magnetic moments on equivalent Mn sites were constrained to be
equal.
\begin{figure}[!h]
\includegraphics[scale=1.0]{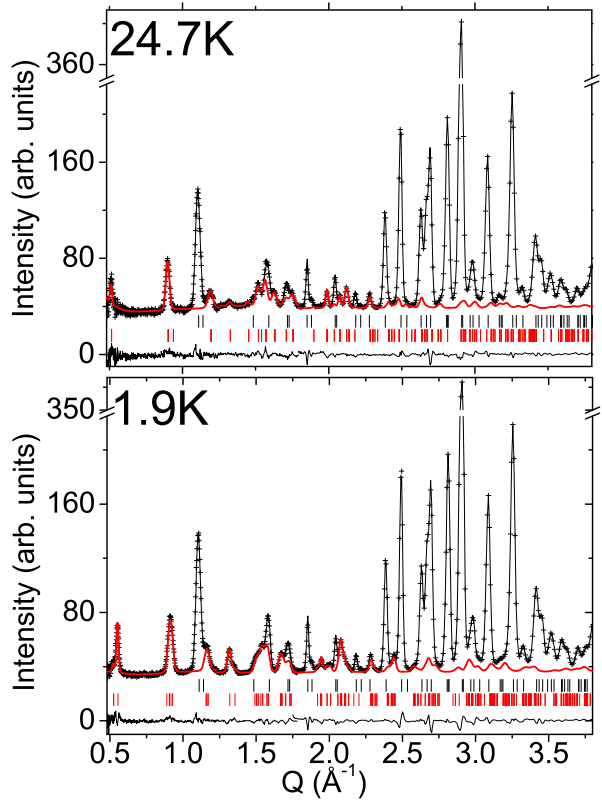}
\caption{Rietveld refinements at 24.7K and 1.9K. Data sets from 3
detector banks located at 18, 35 and 63.6$^{\circ}$ 2$\theta$ are
merged on the same scale. The cross points and solid lines show the
experimental data points and calculated profile respectively. The
difference is shown on the bottom as solid lines. The upper and
lower rows of markers indicate respectively the positions of the
nuclear and magnetic reflections. The red line emphasizes the
calculated magnetic contribution (color online).}
\end{figure}

The crystal structure of YMn$_2$O$_5$ refined at 53K, space group
\textit{Pbam}, is in perfect agreement with previous studies on
isostructural RMn$_2$O$_5$, (R=Tb,Ho,Dy)\cite{blake:214402}. As for
other members of the series, there is no evidence of crystal
symmetry breaking from neutron diffraction at low-temperature, due
to the extremely weak atomic displacements involved. Below 45K, the
data show the appearance of magnetic Bragg peaks, which can be
indexed with a CM propagation vector \textbf{k}=(1/2,0,1/4) above
23K, and with an ICM \textbf{k}$\sim$(0.48,0,0.29) below 17K. The
transition between CM and ICM states, marked by the coexistence of
both magnetic phases, is of first-order type. A small discontinuous
reduction of the magnetic signal is also observed on warming through
the ICM-CM transition. The combined structural/magnetic Rietveld
refinements, shown in figure 1 for data at 24.7K and 1.9K, are of
very good quality, with magnetic reliability factors of 5.6\% and
5.5\% and $\chi^2$ of 4.4 and 16.3 (the higher value is due to the
longer collection time). The corresponding magnetic structures are
displayed in figure 2, and complete lists of parameters are reported
in Table I. In the CM phase, the spin arrangement, corresponding to
an amplitude modulation, is similar to that found for
TbMn$_{2}$O$_{5}$\cite{chapon:177402}. The spins, directed in the
\textit{ab}-plane, are antiferromagnetically aligned
along...Mn$^{4+}$...Mn$^{3+}$Mn$^{3+}$...Mn$^{4+}$... chains running
along the \textit{a}-axis. We note that, within a chain, each
Mn$^{3+}$ ion is connected to two Mn$^{4+}$ located in layers at
z$\sim$0.25 and (1-z), an important detail (see below) that is
difficult to represent in the projected structure. There are two AFM
"chains" per unit cell \cite{chapon:177402}. The moments on the
Mn$^{3+}$ and Mn$^{4+}$ sites are oriented at respective angles of
10(5)$^{\circ}$ and 15(5)$^{\circ}$) to the \textit{a}-axis. The
spins on Mn$^{4+}$ sites at z$\sim$0.25 and (1-z), connected through
the Mn$^{3+}$ layer, are ferromagnetically (FM) aligned. A magnetic
structure with constant moments, as shown in figure 2a, can be
obtained when the phases are set to $\pm \pi /4$ (see below and also
Table I).

The ICM magnetic structure at 1.9K also corresponds to a sinusoidal
modulation of the moments (figure 2). However, this phase is a true
SDW, since every amplitude value is realized on each
crystallographic site. An unconstrained refinement of the initial
model found by Simulated Annealing shows that magnetic moments on
sites related by the glide plane operation are phased by values
close to $\frac{1}{2}$k$_x$ (k$_x$=0.479 being the component of the
propagation vector along a*). This leads to an almost exact
cancelation of the magnetic moment in one of the chains when the
moments in the other chain are fully ordered. The refinement also
indicates that Mn$^{4+}$ atoms at z$\sim$0.25 and (1-z) positions,
unrelated by symmetry operations of the group of the propagation
vector, have phase shifts of almost $\frac{1}{2}$$k_z$, with
k$_z$=0.291. The moments in one of the chains are at an angle of
15(4)$^{\circ}$ to the \textit{a}-axis, similar to that observed in
the CM phase, while the direction of the moments in the other chain
is tilted by 56(4)$^{\circ}$. This model implies the superposition
of the two irreducible representations (irreps) of the paramagnetic
group, as allowed in the case of a first-order transition. The
observed canting is consistent with magnetic susceptibility
data\cite{Kagomiya2}: the CM phase has \textit{b} and \textit{c} as
almost equal "easy" magnetic directions.  At the CM-ICM transition,
\textit{b} becomes a harder axis while \textit{a} becomes an easier
axis, again in agreement with our magnetic structure showing a
rotation of half of the chains towards the \textit{b}-axis. There
are no anomalies in the susceptibility along c at the CM-ICM
transition.

\begin{figure}[h!]
\includegraphics[scale=0.5]{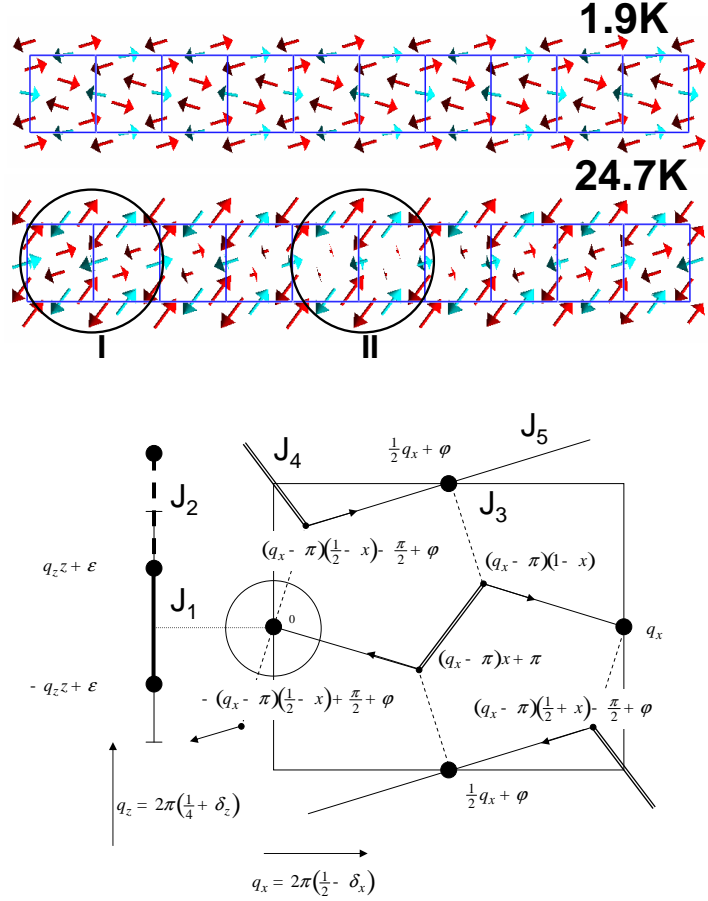}
\caption{(a) Magnetic structure at 24.7K (top) and 1.9K (bottom)
projected in the \textit{ab}-plane. Ten unit cells are displayed
along \textit{a}. For clarity, a single Mn$^{4+}$/Mn$^{3+}$ layer is
shown. The light blue and red arrows represent the spins on
Mn$^{4+}$ and Mn$^{3+}$, respectively. For the ICM structure, region
I is locally very similar to the CM phase (both chains having
sizeable moments), and does not contain inversion centers.  Region
II could potentially contain an inversion center, and is described
in more detail in Figure 2b and in the text. (b) A schematic
representation of the magnetic structures of both CM and ICM phases
(see text). The fragment on the left side represents a portion of
the ...Mn$^{4+}$-Mn$^{4+}$... chains along the \textit{c}-axis. The
SDW phases are as shown in the labels of the Mn$^{3+}$ sites, and
are obtained for the Mn$^{4+}$ sites by adding the values of the
\textit{a}- and \textit{c}-axis projections. The arrows indicate the
direction of the underlying \emph{centrosymmetric} vector field that
is coupled to the magnetism, and coincides with the axes of the
Mn$^{3+}$O$_{5}$ pyramids.  Magnetic exchange pathways are also
indicated (color online).}
\end{figure}

A unified description of both CM and ICM structures, consistent with
the experimental data within the errors, is shown in Fig 2b. Phases
on each crystallographic site within a chain have been assigned so
that the moments follow a single harmonic modulation (their
amplitude being related to their \textit{x} fractional coordinate).
The phases of the waves on adjacent chains are allowed to vary from
being exactly opposite, the global phase shift between them being
denoted as $\varphi$. Along \textit{c}, the moments also follow a
sinusoidal modulation (their amplitude being related to
\textit{z'}), which is the fractional coordinate
(z'=\textit{z}-$\frac{1}{2}$), with a phase shift $\epsilon$ with
respect to the origin. We have deliberately chosen the origin of the
plot to coincide with an inversion symmetry point of the ICM
modulation along the \textit{a} axis for $\varphi$=0 and
$\epsilon$=0. The CM constant-moment phase is obtained by setting
the incommensurability parameters $\delta_{x}$ and $\delta_{z}$ and
the phase shift $\epsilon$ to zero, and by setting $\varphi$=$\pi$/2
(Figure 2b). It is clear by construction that the ICM phase is
non-centrosymmetric for $\epsilon \neq 0$ and $\varphi\neq$ 0.  The
relationship between the CM and ICM phases is also illustrated in
Fig 2a.  The CM phase corresponds to a local region of the ICM phase
that does not contain centers of symmetry.

The value of the \emph{net} electrical polarization along the
\textit{b}-axis (i.e., averaging over the oscillating components) is
easily calculated based on the contribution of the symmetric
inter-chain exchange interaction to the magneto-elastic coupling
described in \cite{chapon:177402} and the phase factors of Fig 2b:

\begin{equation}
P^{ICM}=4C\overrightarrow{S_3}\cdot\overrightarrow{S_4}cos(2\pi(\frac{1}{4}+\delta_z)z')cos(2\pi\delta_x(\frac{1}{2}-x))cos(\epsilon)sin(\varphi)
\end{equation}

where $\overrightarrow{S_3}$ and $\overrightarrow{S_4}$ are the
magnetic moments on the Mn$^{3+}$ and Mn$^{4+}$, respectively, and
$C$ is the magneto-elastic coupling constant. The polarization has
been obtained by multiplying the magnetic terms by the
\textit{b}-axis component of the underlying polar field, shown with
arrows in Figure 2b. From equation (1), it is clear that a non-zero
value of the phase $\varphi$ is required to promote a
\textit{b}-axis polarization. This is only possible if both irreps
are involved, since for each irrep $\varphi$=0 by symmetry. In
addition, the polarization direction can change depending on the
value of $\varphi$.

With the help of eq. (1) and the experimental values of the phases
(Table I), we can predict the value of the polarization for the ICM
phase. In practice however, the error bars on the magnetic phases
for an individual measurement introduce a large uncertainty on the
value of the calculated polarization. To overcome this, we have
fitted the weak temperature dependence of the ICM phases, to obtain
constant phase differences. These yield a polarization of
\emph{opposite sign} and reduced by approximately a factor of 5 with
respect to the CM phase, in close agreement with the experimental
values determined by Kagomiya \textit{et al.} \cite{Kagomiya2}. The
temperature dependence of the spontaneous polarization has been
calculated, and is shown in Figure 3 to be in good agreement with
the experimental curve by Kagomiya and co-workers \cite{Kagomiya2}.

\indent Although $\varphi\neq$ 0 is clearly allowed by symmetry in
our case, and as we have shown this can lead to a spontaneous
polarization of either sign in the ICM phase, it is presently
unclear to us how a non-zero value can be energetically favorable.
In fact, in the simple isotropic exchange model, the inter-chain
energy can be written as:\\

\begin{equation}
E^{ICM}_{3}= -4J_{3}\overrightarrow{S_3} \cdot \overrightarrow{S_4}cos(2\pi(\frac{1}{4}+\delta_z)z')sin(2\pi\delta_x(\frac{1}{2}-x))cos(\epsilon)cos(\varphi)\\
\end{equation}

which is even in $\varphi$. It is noteworthy however, that
$E^{ICM}_{3}$ is linear in $\delta_{x}$ for small $\delta_{x}$,
suggesting a natural mechanism to stabilize the incommensurability
along the \textit{a} axis.

\begin{figure}[h!]
\includegraphics[scale=1.0]{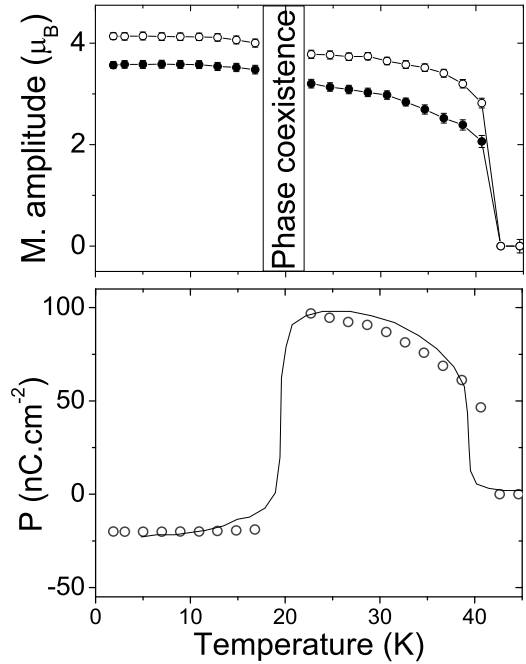}
\caption{(a) Refined values of the magnetic wave amplitudes on
Mn$^{3+}$ (open symbols) and Mn$^{4+}$ (filled symbols) as a
function of temperature.  The \textit{average} moment on each site
is $\frac{1}{\sqrt{2}}$ of the wave amplitude. (b) \textbf{Symbols}:
electrical polarization of YMn$_2$O$_5$, as calculated from eq. (1).
\textbf{Solid line}:  Experimental values of the electrical
polarization, extracted from Kagomiya \textit{et al.}
\cite{Kagomiya2}. The calculated polarization has been scaled by a
single constant to account for the unknown magneto-elastic coupling
parameter.}
\end{figure}

In summary, we have developed a model to explain the presence of
ferroelectricity in the CM and ICM phases of YMn$_2$O$_5$
\emph{without} the need to invoke non-collinearity and/or chirality
as essential features.  We show that ferroelectricity is compatible
with an acentric SDW, which the literature reports as the most
probable magnetic structure for this class of
materials\cite{Wilkinson,gardner}. We have determined the
high-temperature commensurate and low-temperature incommensurate
magnetic structures of YMn$_2$O$_5$ based on neutron diffraction
data. The calculated electrical polarization based on our model and
a simple magneto-elastic coupling was found to be in good agreement
with the experimental values at all temperatures, including a
previously unexplained sign reversal at the CM-ICM transition.
\indent We would like to acknowledge helpful discussions with Daniel
Khomskii, Maxim Mostovoy and Joseph Betouras.

\begin{table} \caption{\label{tab:table1}Magnetic parameters obtained
from Rietveld refinements of the 1.9K and 24.7K data. }
\begin{ruledtabular}
\begin{tabular}{ll|lll|lll}
 & & & 24.7K & & & 1.9K & \\
Atom & Position & M$_x$($\mu_B$) & M$_y$($\mu_B$) & Phase(rad) &
M$_x$($\mu_B$) & M$_y$($\mu_B$) & Phase(rad) \\
\hline
Mn$^{4+}$ & (0 0.5 0.255) & 3.092(9) & -0.5(2) & $\frac{\pi}{4}$ & -3.46(8) & -0.9(1) & -0.459 \\
Mn$^{4+}$ & (0.5 0 0.255) & -3.092(9) & -0.5(2) & $\frac{\pi}{4}$ & -1.9(2) & -3.0(1) & 4.15(69)\\
Mn$^{4+}$ & (0 0.5 0.745) & 3.092(9) & -0.5(2) & $\frac{\pi}{4}$ & -3.46(8) & -0.9(1) & 0.459 \\
Mn$^{4+}$ & (0.5 0 0.745) & -3.092(9) & -0.5(2) & $\frac{\pi}{4}$ & -1.9(2) & -3.0(1) & 5.09(69) \\
\hline
Mn$^{3+}$ & (0.412 0.351 0.5) & -3.63(9) & -1.0(2) & $\frac{\pi}{4}$ &  -4.02(8) & -1.0(2) &  2.83(44) \\
Mn$^{3+}$ & (0.588 0.649 0.5) & 3.63(9) & 1.0(2) & $\frac{\pi}{4}$ & -4.02(8) & -1.0(2) & -0.31(44) \\
Mn$^{3+}$ & (0.088 0.851 0.5) & 3.63(9) & 1.0(2) & $\frac{\pi}{4}$ & -2.5(2) & -3.3(2) & 1.44(81) \\
Mn$^{3+}$ & (0.912 0.149 0.5) & 3.63(9) & 1.0(2) & $\frac{\pi}{4}$ & -2.5(2) & -3.3(2) & 1.44(81) \\

\end{tabular}
\end{ruledtabular}
\end{table}

\end{document}